# Magnonic band structure in a Co/Pd stripe domain system investigated by Brillouin light scattering and micromagnetic simulations

*Chandrima Banerjee,[1] Pawel Gruszecki,[2] Jaroslaw W. Klos,[2] Olav Hellwig,[3] Maciej Krawczyk[2] and Anjan Barman[1]\**

[1]*Department of Condensed Matter Physics and Material Sciences, S. N. Bose National Centre for Basic Sciences, Kolkata 700 106, India*

[2]*Faculty of Physics, Adam Mickiewicz University in Poznan, Umultowska 85, Poznan, 61-614, Poland*

[3] *Institute of Physics, Chemnitz University of Technology, Reichenhainer Straße 70, D-09107 Chemnitz, Germany and Institute of Ion Beam Physics and Materials Research, Helmholtz-Zentrum Dresden–Rossendorf, 01328 Dresden, Germany*

*\*Email: abarman@bose.res.in*

*Abstract*

By combining Brillouin Light Scattering and micromagnetic simulations we studied the spin-wave dynamics of a Co/Pd thin film multilayer, features a stripe domain structure at remanence. The periodic up and down domains are separated by cork-screw type domain walls. The existence of these domains causes a scattering of the otherwise bulk and surface spin-wave modes, which form mode families, similar to a one dimensional magnonic crystal. The dispersion relation and mode profiles of spin waves are measured for transferred wave vector parallel and perpendicular to the domain axis.

The possibility to use spin-waves (SW) to excite, transmit, store, and retrieve electric signals as well as to perform logical operations has fueled a new spectrum of research in the wave-based signal processing technology [1-4]. Usually, magnonic crystals (MC), *i.e.* arrays of macroscopic magnetic stripes [5], dots [6], antidots [7], etched grooves or pits[1], periodic variations of the internal magnetic field [9] and saturation magnetization by ion implantation [10], are employed to control the flow of SWs. Magnetostatic surface SWs or the Damon–Eshbach (DE) SW modes, which propagate perpendicular to the in-plane (IP) magnetization direction, are promising in this context because of their large group velocities ($v_g$) and low attenuation. Consequently, significant progress has been made towards the practical realization of magnonic devices in terms of the on-chip generation, directional channeling, detection and manipulation of SWs. However, to realize the experimental geometry, a large magnetic field has to be applied to enforce the magnetization perpendicular to the SW propagation direction. This is a major obstacle to the implementation of a MC into a practical device. A way out is to use the Oersted field generated from an underlying current-carrying stripe[11-12], which is still plagued by the problem of generation of waste heat which increases with increasing data processing speed. Moreover, the fabrication of periodic nanostructures involves high-precision electron-beam lithography which is very complex and expensive. An alternative approach to overcome these fundamental drawbacks is to take recourse to the spin dynamics in the remanent state, which is more suitable for nanoscale device applications, as it does not require any stand-by power once initialized.

In the literature, only few reports exist on the magnetization dynamics for systems with an inhomogeneous magnetization distribution containing domains and domain walls. Most commonly studied is a periodic or non-periodic distribution of magnetic units of up and down domains (parallel, labyrinthine, bubble-like domains)[13-19], separated by negligibly thin, one-dimensional domain walls. Such domains appear in magnetic thin films with a perpendicular anisotropy smaller than the demagnetizing energy if the film thickness is higher than a critical value $t_c$, which is, in turn, dependent on the perpendicular anisotropy constant, saturation magnetization and the exchange constant. The morphology of the domains and consequently the spin dynamics are strongly affected by the field history, allowing for a broad range of tunability. Moreover, just recently the topic of SWs propagation within domain walls, as an ultra-narrow waveguides is intensively studied by the magnonic community, however, mostly theoretically in magnetic systems with strong Dzyaloshinskii-Moriya interaction[20-23]. There are only a few examples of experimentally realized systems presenting SW propagation within domain wall [24].

The study on the magnetic properties of stripe domain systems (array of parallel up and down domains) was pioneered by Kittel [25], where the formation of stripe domains was predicted for films thicker than the magnetic domain width. After that, a multitude of theoretical [26-27] and experimental [14, 28-30] studies have been carried out in this direction for a broad variety of magnetic systems, where the evolution, morphology and interaction between the stripe domains are addressed. Experimentally, the formation of a stripe domain structure is characterized as a function of film thickness, applied magnetic field, annealing temperature, film composition, etc., using magnetic force microscopy (MFM), magneto-optical Kerr effect (MOKE), X-ray diffraction, for NiFe [31-32], Co[33-34], CoFeSiB [35], FeSiB [36-37], FeCoAlON[38], FeCoZr[39], FePd[40] films. Nevertheless, only few measurements have been reported so far on the dynamical properties of magnetization, which used ferromagnetic resonance (FMR) [4-6] and Brillouin light scattering (BLS) [14,43] techniques. The results reveal that the consecutive domains give rise to periodically modulated internal magnetic field, which introduces a partial reflection of the propagating SW at the stripe domain boundaries. The feature of rotational magnetic anisotropy is also evidenced, which is associated with the reorientation of the stripes. These results triggered the possibility towards achieving reconfigurable magnetic properties by tailoring the intricate interaction between the consecutive stripes as well as the magnetocrystalline anisotropy in these systems.

In this present work, we aim at developing a combined understanding of the static and dynamic magnetic properties of the parallel stripe domain structure in Co/Pd multilayered systems. The study of magnetic multilayers with high perpendicular magnetic anisotropy (PMA) is motivated to a large extent by industrial importance of these films, *e.g.*, in patterned magnetic media [44], spin transfer torque magnetoresistive random access memory (STT-MRAM) [45] and magnonic crystals [46]. In these systems, the interface anisotropy (because of the broken symmetry), combined with *d-d* hybridization at the ferromagnet/nonmagnet interfaces, are responsible for the onset of PMA, and consequently of a stripe domain structure, for film thickness greater than a critical value. Moreover, the proximity effect, along with the magnetic coupling between the layers leads to magnetic properties unreachable to standard single-layer systems. These effects get further modified in presence of stripe domain. In this context, a deep understanding of the magnetic coupling between the constitutive layers at larger stack thickness and the impact of stripe domain on the static and dynamic properties become of crucial interest. So far, the quasistatic properties of the domain structure have been investigated for FeNi/Co [47], FePd[48][7], Fe/Cu [49], Co/Cu/(Fe/Ni)/Cu(001) [50]and Co/Au [51] multilayers. However, the dynamic SW properties are still unexplored.

Here we have investigated the static and dynamic magnetic properties in Co/Pd multilayers with stripe domains by using MFM, vibrating sample magnetometry (VSM) and Brillouin light scattering (BLS) spectroscopy. Our first goal is to examine how this stable spin configuration manifests itself on the dynamic response. For that, the SWs were measured in saturated as well as in remanent state (*i.e.* in presence of stripe domain) and their wave-vector dispersions are analyzed and compared. We have further studied the evolution of the SW band structure by changing the relative orientation between the SW wave-vector and the domains. Micromagnetic simulations are employed to calculate both eigenvalues (mode frequencies) and eigenvectors (mode profiles) of the SWs. The results proclaim that the SW propagation is similar to that in a one-dimensional (1D) magnonic crystal, which can be significantly reconfigured by varying the wave vector orientation. Furthermore, this system serves as a possible prototype for magnon-based data transfer devices, with the advantage of non-volatile operation, energy efficient computing, ease of fabrication and control on SW frequency and velocity.

**Experimental Details**

The multilayered sample was deposited using DC magnetron sputtering and has the following structure: Ta(15)/Pd(30)/[Co(10)/Pd(7)]$_{25}$/Pd(20), where the numbers denote the thickness in Å. For deposition we used a multi-source confocal sputter up geometry with a source/substrate distance of 4–6 inches. During deposition, the substrate was rotated for uniformity at about 3 Hz and was kept at room temperature using a low Argon pressure of 3mTorr. Co was sputtered at 250W with a deposition rate of 0.7 Å/sec and Pd was sputtered at 100 W with a deposition rate of ~2 Å/sec. Given the low Ar pressure used during deposition, the film is relatively smooth, with sharply defined interfaces [52].

To obtain the aligned stripe domain structure, the saturating magnetic field was applied within the film plane (arbitrary direction, as the films are [111] out-of-plane (OOP) textured and IP isotropic) followed by an IP AC demagnetization (preferred alignment of the domains is along the applied field direction), *i.e.* the IP field is switched back and forth between positive and negative polarity starting above the saturation field strength down to about a few Oersted, where the field amplitude is reduced about 1% in each step. As a result, we obtain an aligned parallel magnetic stripe domain state similar to an array of artificial stripe wires, *i.e.* a magnonic crystal. The magnetization curves (IP and perpendicular to the plane) were measured using VSM at room temperature. The domain images at the remanent state were recorded by magnetic force microscopy (MFM). BLS measurements in the 180° backscattered geometry were employed to investigate the wave-vector dispersions of the thermal magnons. This technique relies on an inelastic light scattering process due to interaction between incident photons and magnons. Monochromatic laser light (wavelength $\lambda = 532$ nm) from a solid state laser was focused on the sample surface. As the laser beam is inelastically scattered from the magnons, due to conservation of momentum, the magnitude of the IP transferred wave-vector $q$ depends on the incidence angle of light $\theta$ according to: $q = (4\pi/\lambda)\sin\theta$. The direction of magnon wave-vector $q$ lies along the intersection of the scattering plane and the film plane. Cross polarizations between the incident and the scattered beams were adopted in order to minimize the phonon contribution to the scattered light. Subsequently, the frequencies of the scattered light are analyzed using a Sandercock-type six-pass tandem Fabry-Perot interferometer (JRS Scientific Instruments) [53]. One set of measurements was carried out in the demagnetized state in presence of the stripe domains for $q$ parallel and perpendicular to the stripe axis. This was accomplished by rotating the sample around the film normal, *i.e.* varying the angle $\varphi$ between $q$ and stripe axis. In another measurement set, the sample was saturated applying a strong IP magnetic field ($H = 3.5$ kOe) and the wave-vector dispersion was taken in the DE geometry ($q \perp H$).

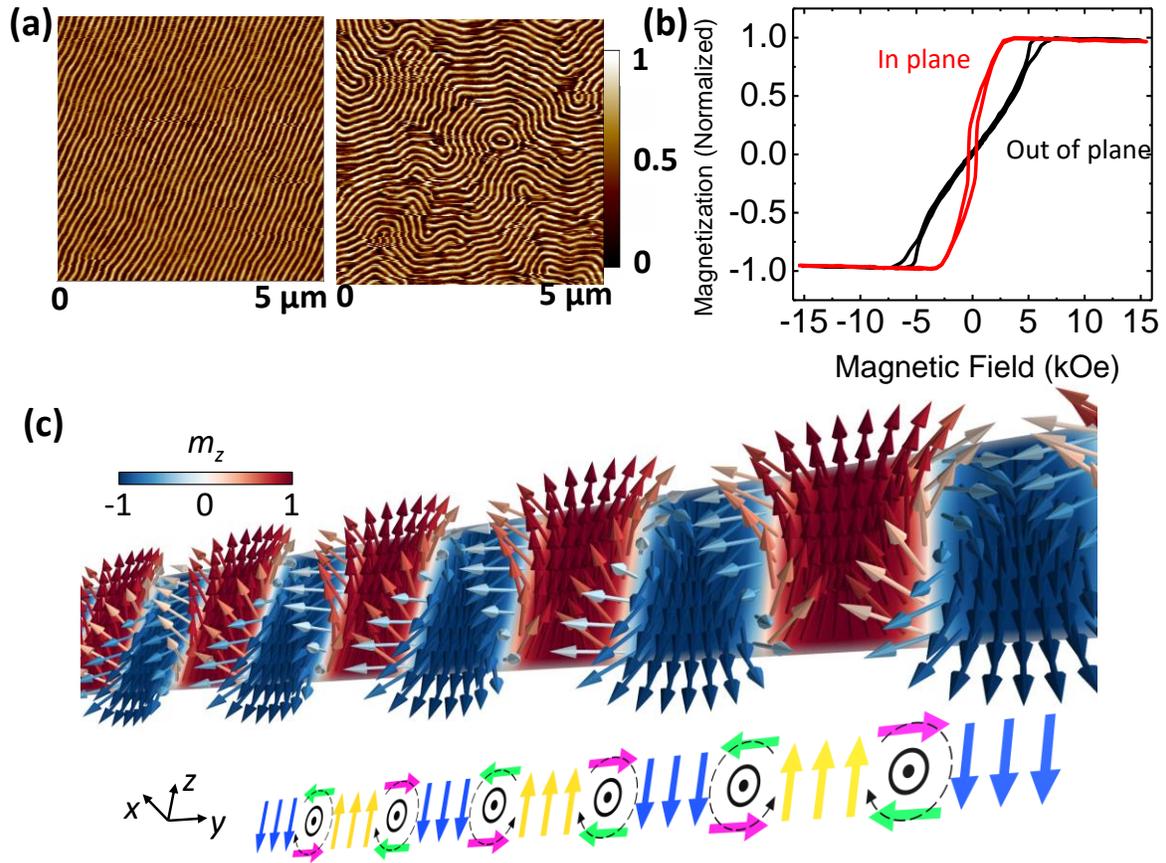

**Figure 1:** (a) MFM image in the demagnetized state showing the parallel (left panel) and labyrinth (right panel) stripe domains of a Co/Pd multilayered system, obtained after IP and OOP AC demagnetization, respectively. (b) Magnetic hysteresis loops of the multilayers with IP and OOP applied magnetic fields. (c) The domain configuration (cross-sectional view) obtained from micromagnetic simulations. Arrows correspond to the magnetization vector and colors spanning between blue and red correspond to the OOP component of the magnetization. The bottom panel schematically visualize static magnetization with clockwise and anticlockwise domain walls. The $x$ and $z$ axes are assumed to be parallel to domain walls and multilayered system thickness, respectively.

**Micromagnetic simulations**

To validate the experimental results and to understand the physical character of the resonant modes we exploited micromagnetic simulations using MuMax3 [53]. Micromagnetic simulations primarily were used to relax the stable stripe domain pattern of comparable lattice constant as obtained experimentally, to compute dispersion relations for the uniformly IP magnetized structure and the stripe domain pattern, and, finally, to visualize resonant mode profiles. In micromagnetic simulations we assumed effective homogeneous material parameters in the whole structure. The saturation magnetization and anisotropy constant were set to the values obtained from the VSM measurements. To fit micromagnetic simulations results with the BLS data, we had to assume an anisotropic exchange constant, [54-56] with $A_{ex,IP}=1.45\times10^{-6}$ erg/cm and $A_{ex,OOP}=1.05\times10^{-6}$ erg/cm along the IP and OOP directions, respectively. Micromagnetic simulations were performed for the system of thickness $L_z=42.5$ nm discretized with the unit cell of dimensions being much smaller than the exchange length or the width of domain walls. The SW dispersion relation and resonant modes profiles were calculated according to Ref. [57]. Obtained results are presented in a form of color maps where color intensities correspond to the OOP component of the magnetization and relate to the intensities obtained in measurements (Fig. 2(g-i)). Resonant mode profiles (Fig. 3) are presented as a distribution of the amplitude and phase of the OOP dynamic magnetization component ($m_z$), a component to which BLS is sensitive to. A detailed description of micromagnetic simulations is provided in the supplementary materials.

**Results and Discussions**

The stripe domain patterns as measured by MFM in the demagnetized states are shown in Fig. 1a. After IP AC demagnetization, we observe well-defined parallel stripe domains aligned along the IP demagnetization direction. The half period of the pattern is $A/2\approx60$nm ($A$ being the lattice constant), which is comparable to the thickness of the entire stack (42.5 nm). As mentioned in the introduction, the broken translational symmetry at the Co/Pd interfaces, combined with the *d-d* hybridization is responsible for the PMA in this structure, which favors an OOP magnetization. The tendency of the magnetization to rotate OOP, however, is hindered by the presence of the magnetic charges on the surface of the film, which contributes to the magnetostatic energy (shape anisotropy). The competition between the PMA and the thin-film shape anisotropy leads to domains with up and down magnetization perpendicular to the film plane [30]. The features of stripe domain pattern depend on the respective thicknesses of the layers, the number of repetitions of the Co/Pd bilayer, and the amplitude of the anisotropy field. Basically, the magnetic domains occur when the gain in magnetostatic energy due to the domain structure is bigger than the energy required to form the domain walls. A quality factor is defined, as given by, $Q = \dfrac{K_u}{2\pi M_S^2}$, where $K_u$ is OOP anisotropy energy density while $M_S$ is saturation magnetization. For a moderate value of the quality ratio, $Q \leq 1$, a ground state with a stripe domain structure is favored.

The static magnetometry results for applied field along IP and OOP directions are shown in Fig. 1b. The higher saturating field (~7.5 kOe) observed on the OOP hysteresis loop suggests that the magnetization is primarily in the sample plane. The IP hysteresis loop marks two distinct magnetization phases: an IP component, which quickly reverses at fields close to the coercive field, and a linear approach to saturation (starting at around $H\approx1500$ Oe), due to the perpendicularly magnetized stripe domains, whose magnetization progressively rotates under the application of an IP magnetic field. The coercivity is approximately 0.7 kOe and the remanent magnetization is around 30%, revealing that the magnetization vector is tilted *w.r.t.* the OOP direction. The magnetization reversal curve for OOP magnetic field, on the other hand, is characterized by domain nucleation, propagation, and annihilation characteristics of labyrinth domain structure [15]. As discussed in Ref. 15, the parallel domains are robust and stable under small OOP field exposures up to about +/- 3-4 kOe, *i.e.*, within the linear region of the OOP hysteresis loop. In this range, only the relative width of up versus down domains is altered by the out-of-plane field although the overall domain topology is preserved. Figure 1a (right panel) shows the MFM image taken after demagnetizing the film with perpendicular applied field, which confirms the labyrinth domain formation. The saturation magnetization and the uniaxial anisotropy constant associated with the PMA were both estimated from the hysteresis curves as $M_s$=970 emu/cc, and $K_u$=1.9×10$^6$ erg/cm$^3$ [58]. These values give a quality factor $Q$=0.3.

We first describe the BLS measurements of the wave-vector dispersion of thermal SWs after saturating the sample magnetization by applying an IP field $H$=3.5 kOe. The measurement geometry is shown in Fig. 2a, where the transferred IP wave-vector is perpendicular to direction of magnetization, *i.e.* DE geometry. Figure 2g shows measured SW frequency peaks (solid symbols, labelled with M) as a function of $q$, together with a typical spectrum taken at $q$=1.2×10$^7$ rad/m (Fig. 2d). We observed two dispersive modes where the mode M1 (M2) shows a maximum (minimum) at $q_g$=1×10$^7$ rad/m, leading to a frequency gap of 3 GHz.

To validate the experimental results and to understand the physical character of the resonant modes we exploited micromagnetic simulations. The simulated mode frequency in dependence on the wave-vector is presented by the gray scale map in Fig. 2g (marked with label S), which agree well with the experimental results. Further, we investigated the physical nature of the modes. Figure 3a displays the corresponding power and phase maps of the SW modes at $q$=2×10$^7$ rad/m. The result reveals the mode M1 (S1) extends across the volume of the sample, however with increasing localization of the amplitude at the bottom surface of the film as the wavenumber increases. This points at the DE character of the mode, although with non-monotonic dispersion. The former is also accorded by the marked Stokes/anti-Stokes asymmetry of the mode in the BLS spectrum, which can be reversed by simply reversing the direction of the magnetic field. We attribute this unusual non-monotonicity to the PMA, as confirmed with the analytical model [59], which reproduces the BLS as well as the micromagnetic simulation dispersion relations. The mode M2 (S2) is a perpendicular standing spin wave (PSSW) excitation of the exchange character with the one nodal plane in the middle of the film (Fig. 3a).

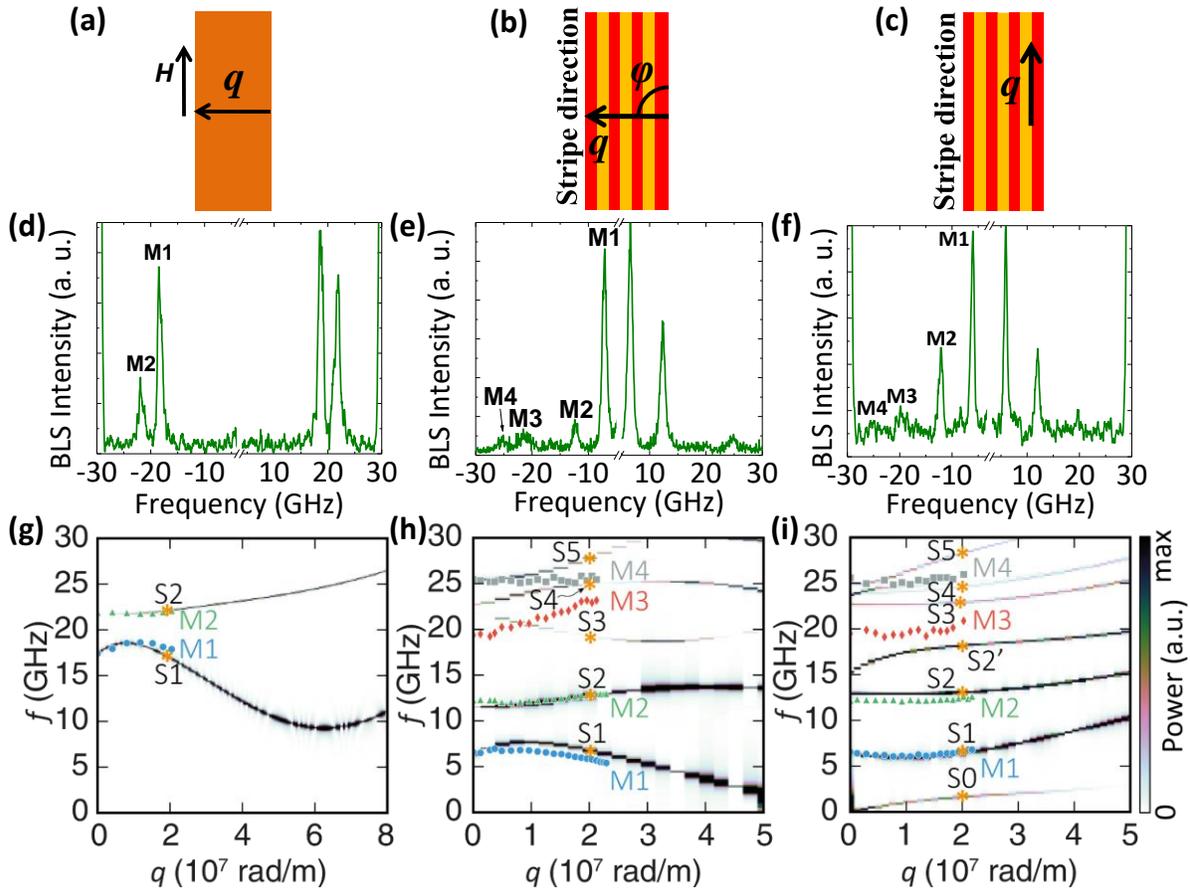

**Figure 2:** (a)-(c) The measurement geometries for the saturated state, $\varphi=90°$ in (a) and (b) and $\varphi=0°$ in (c). (d)-(f) Corresponding typical BLS spectra at $q=1.2\times10^7$ rad/m and (g)-(i) dispersion curves. Measured in BLS dispersions and modes are labeled with M and marked with the color points in (g)-(i). Simulation results are shown by the gray scale map, the orange stars labeled with S represent modes visualized in Fig. 3.

In the next step, we investigated how the SW dynamics are modified in presence of a stripe domain structure. Figure 1c presents the cross-sectional view of the simulated stripe domain structure where the magnetic state is shown with the arrows and OOP component of the magnetization by a red-white-blue color map and arrows' color. Clearly, it consists of basic up and down domains, with a spatial periodicity of about $A$=100 nm which is in reasonable agreement (slightly underestimated) with the experimental MFM images. These domains are separated by complex twisted domain walls (DW), as given by the not fully vertical arrows of colors different than saturated dark red and dark blue being present in the middle part of each domain, where the magnetization is exactly parallel to the $z$-axis.

Note that, the DW occupies a significant region in each spatial period, so that the magnetization profile in each domain deviates from its rectangular shape. For the DWs, interestingly, two main features are evidenced as follows. i) The magnetization profile in the DW depicts a Neel-wall like character at the film surfaces, which eventually rotates towards the direction parallel to the DW axes when moved away from the surface (Bloch-wall like) [see schematic representation of domain structure in Fig. 1c (lower panel)]. ii) At the same time, it evolves into a twisted pattern around the DW axis (cork-screw type) [60], which is clockwise or anti-clockwise for alternate DWs (see Fig. 1c). Consequently, the resultant magnetization inside the DW is directed IP along the DW axis. This spin configuration is different from the closure domain structure reported earlier for single layered films. In fact, the spin structure is governed by the minimization of total energy and as reported in Ref. [27], the dipolar interaction between consecutive domains is minimal when the magnetization vector stays in the DW plane. This is easily conceivable because the IP magnetic component of the wall should be in the same direction as the magnetic field that was applied during the formation of the stripes, which means in the IP AC demagnetization process we align the IP components of the domain walls rather than the perpendicular domains themselves.

In Figs. 2h and i, we have plotted the resonant frequencies versus $q$ at $H$=0, for $q$ perpendicular ($\varphi$=90°) and parallel ($\varphi$=0°) to the stripe domain axis, respectively. The solid symbols represent the measured values, while the simulated frequencies are shown by the color map. The typical BLS spectra (Figs. 2e and 2f) are characterized by the presence of four peaks, however, with different frequency versus wavenumber dependences for the orthogonal direction of $q$, which merge to the same frequencies as $q$ approaches 0. The increase in the number of peaks as compared to the saturated state signals results from the periodical spin texture. In case of SWs propagation across DWs those magnetization non-uniformities are treated as regular scattering centers, which are provided by the domain boundaries, similar to the geometric magnetic boundaries of a lithographically patterned magnonic crystal. In case of SWs propagating parallel to the DWs those inhomogeneities are treated as different channels defined by different magnetization orientation. In both cases periodicity is crucial to collect strong enough BLS signal from the SW excitations. That is confirmed by BLS measurements of a labyrinth-like domain structure, where the recorded spectra are very noisy and almost wave-vector independent (See supplementary materials).

For $\varphi=90°$, the corresponding SW frequencies exhibit pronounced dispersion with $q$, together with characteristic band gaps (Fig. 2h). The width of the lower band gap is larger than the frequency gap observed for the saturated state (see Fig. 2g). Good agreement between experimental and simulation results is visible for M1 (S1) and M2 (S2), whereas the frequency of S3 in simulations is shifted up with respect to BLS result M3. Simulations show also, that the horizontal dispersion line M4, being related to the mode with two nodal lines across the thickness S4 (see Fig. 3b), is hybridized with the other dispersive mode S5. In BLS measurements there is no indication for the hybridization. To explain those discrepancies we note that the lattice constant of the stripe domains in micromagnetic simulations is 100 nm while from MFM images we got around $A=120$ nm. Therefore, in the sample the Brillouin zone boundary is at smaller wavenumber than in micromagnetic simulations. It means that M3 band, being two times folded-back M2 band to the 1$^{st}$ Brillouin zone, should be shifted down with respect to the frequency obtained in the simulations. Moreover, its next folding at $q=0$ should hybridize with the horizontal M4 band for much large wavenumbers being above measured range. The analysis of the modes (see Fig. 3b) supports this explanation, modes S2 and S3, have the same origin in PSSW with one nodal line, whereas horizontal S4 is the first band from another family of PSSWs with two nodal lines across the thickness.

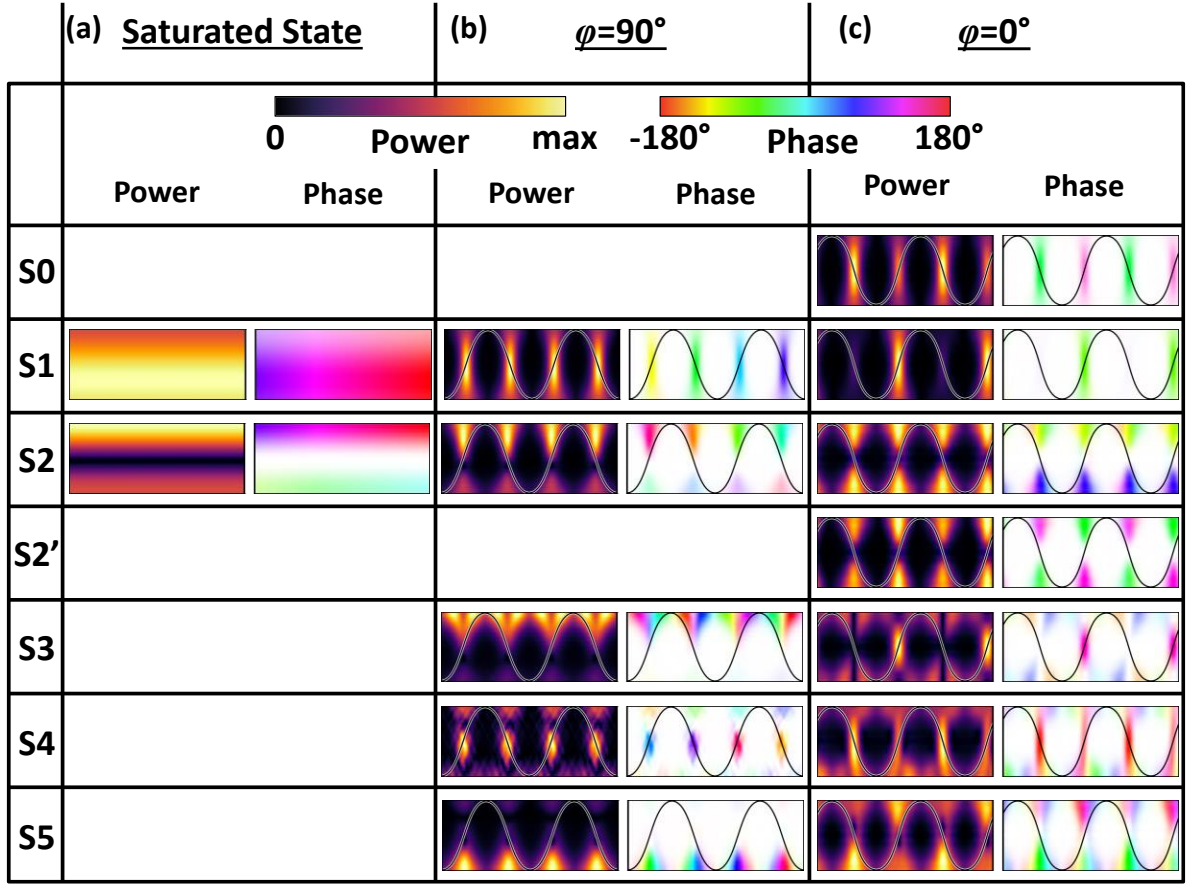

**Figure 3:** The amplitude and phase maps of $m_z$ for different SW modes (as marked with orange stars in Fig. 2) calculated at $q=2 \times 10^7$ rad/m for the saturated state (a), and domain structure with $\varphi=90°$ (b) and $\varphi=0°$ (c). The corresponding color scale is shown at the top of the figure. Dark solid lines in (b) and (c) correspond to the OOP component of the static magnetization at the top surface of the structure.

Interestingly, in Fig. 2h the periodicity of the static magnetization configuration $\pi/A$ ($\approx 3.14 \times 10^7$ radm$^{-1}$) is not visible in the dispersion, while the period $2\pi/A$ becomes clear, whenever the dispersion is plotted in the extended range of $q$ (see supplementary materials). This means that for the $z$ component of the SW amplitude the periodicity is determined by the size of a single domain. However, if we introduce into the analysis an $m_x$ component of the magnetization (parallel to the domain walls) the periodicity of the dispersion relation will be determined by $\pi/A$ (see supplementary materials). Moreover, there appeared dispersion line at low frequency (S0 for $\varphi=0°$, see also Fig. 2i) with the linear slope approaching zero-frequency when $q$ goes to 0 and the S1 branch for $\varphi=90°$ (Fig. 2h) decreasing to 0 at $q=2\pi/A$. This points at that this is a Goldstone mode connected with the DW oscillations [61] which is further confirmed by the dynamic magnetization profiles (see supplementary materials). This also points at possible interactions between DW oscillations and SWs.

On the other hand, for $\varphi=0°$, the dispersion is repressed and all the modes exhibit similar frequency evolution–see Fig.2i. Very good agreement between BLS (M1, M2) and simulation (S1 and S2) results for the first two bands is found, also the fourth band (M4 with S4) match well, apart from the small wavenumbers. The spatial distribution of the modulus of the dynamic magnetization, as will be discussed in the next paragraph, illustrates that the modes primarily reside in the DWs and at the top and bottom surfaces of the structure. This implies that in absence of external magnetic fields, the IP magnetostatic field trapped in the direction of the DW axis actually acts as the driving field for the SW propagation. In that sense, the $\varphi=90°$ and $\varphi=0°$ cases can effectively be termed as the DE geometry and the backward volume magnetostatic SW geometry, although they have different physical origin than in homogeneously magnetized films.

The nature of the above SW excitations can be better understood by carefully looking at the relevant cross-sectional spin precession profiles, which are depicted in Figs. 3b and c, for the two experimental geometries shown in Figs. 2b and c, respectively. The solid black lines represent static $m_z$ component at the top surface of the domain structure indicating the up and down domains for reference. For $\varphi=90°$ geometry (Fig. 3b), the amplitude of the SW modes are localized mainly in the DWs. This confirms that the SW modes investigated in BLS are primarily related to the IP component of the static magnetization. In particular, the mode profiles are characterized by one or more distinct nodal planes, which are indeed close to those reported earlier for 1D MC, prepared by periodic patterning of magnetic material [62-63]. The mode profiles for $\varphi=0°$ are similar to $\varphi=90°$, except that the amplitude is concentrated in alternate DWs, typical of backward volume modes in MCs [64]. These facts prove the equivalence of our stripe domain structure with the previous MCs.

Finally, we elucidate the salient magnetic properties of the Co/Pd multilayered system, as it evolves from the uniformly saturated state to the stripe domain configuration. One sees that the dispersion characteristics, magnonic bandgap width and position are controllable by the external applied field and the orientation of the SW wave-vector. In particular, the evolution of the SW modes and the mode profiles from the uniformly saturated state to the stripe domain state resemble the transformation of magnonic properties from unpatterned continuous thin films to a periodically patterned 1D MC. Essentially, the periodicity of the IP magnetization (inside the DW) promotes a periodic potential, which subsequently leads to a scattering of the SWs. Overall, once the domain structure is formed, this system does not require any external power for SW propagation and thereby can serve as a building block for the practical realization of energy efficient magnonic devices operating in the microwave frequency range.

**Conclusions**

In conclusion, the quasistatic and dynamic magnetic properties of Co/Pd multilayered films characterized by magnetic stripe domains at remanence have been investigated both experimentally and using micromagnetic simulations. The signature of the stripe domains are clearly observed in the MFM and VSM results. The dynamical magnetic properties were analyzed by studying the wave-vector dispersion of the SWs, by means of BLS technique. In the saturated state, two dispersive SW modes are observed, which are identified as the bulk mode and a surface mode of the system. Demagnetizing the film with an AC IP applied field gives rise to an anisotropic remanent state of parallel (rather than a labyrinth) stripe domains aligned along the demagnetization field direction. Micromagnetic simulations reveal the presence of cork-screw type twisted domain walls, with its resultant magnetization directed along the DW axis. The spatial periodicity of this magnetization serves as a periodic potential for SW propagation, in a similar manner as for a laterally confined geometry, *e.g.*, magnonic crystals. As a result, the number of SW modes increases in the BLS spectra and the dispersion depends strongly on the IP angle $\varphi$ due to the IP anisotropy induced by the parallel alignment of the stripe domains. Analysis of amplitude and phase maps of the resonant modes further illustrates that the dynamics of the system is dominated by the magnetization of periodic DWs, yielding mode profiles close to those of a 1D MC. Due to simple, uniform design, such structures are free from fabrication defects and structural non-uniformities, such as edge roughness. Finally, the observed tunability of the magnonic properties opens a new route towards low energy units for magnonic computing, such as microwave filters, splitters and waveguide structures.

**Acknowledgements**

This work was supported by Department of Science and Technology, Government of India (grant no. SR/NM/NS-09/2011(G)), Indo-Polish join Project DST/INT/POL/P-11/2014, the National

Science Centre Poland grant UMO-2012/07/E/ST3/00538, and the EU's Horizon2020 research and innovation programme under the Marie Sklodowska-Curie GA No. 644348 (MagIC).